# Molecular simulation of melting in tetracosane ($C_{24}H_{50}$) monolayers and bilayers on graphite

Cary L. Pint

Department of Physics, University of Northern Iowa, Cedar Falls, IA  50614
Electronic Mail:  cpint@uni.edu

**ABSTRACT**

This work reports an investigation into the solid phase behavior and melting behavior of tetracosane ($C_{24}H_{50}$) monolayers and bilayers physisorbed onto the graphite basal plane using molecular dynamics simulations performed in the constant ($N,V,T$) ensemble.  This study is conducted in order to perform an in-depth analysis of the melting process in the quasi-2D monolayer, and compare the molecular behavior in the monolayer and the bilayer at temperatures near and at the melting transition.  The bilayer is studied by looking at quantities for the bottom and top layer separately in order to break apart the evolution of differing structural and molecular properties that are normally associated with melting in the monolayer.  Simulations suggest that the melting temperature in the bilayer is increased significantly from that of the monolayer- which already occurs at temperatures higher than the bulk tetracosane melting temperature.  In addition, a layer-by-layer melting effect is observed where significant factors associated with the melting transition- the formation of molecular gauche defects and the perpendicular rolling of the molecule chains- occur in the top layer of the bilayer at significantly lower temperatures than they occur in the bottom layer.  Variations are conducted to better understand the role of different interactions in melting of the crystalline monolayer, and the results of simulations are discussed in comparison to recent STM and diffraction experiment and found to be in good agreement.

## 1 Introduction

One of the fastest growing and most fascinating fields of science today involves the study of how very small films of atoms and molecules behave when they interact with or are confined by other surfaces with significantly different properties than those of the film. In general, one can imagine many scenarios that would be equally interesting, and a great effort today is being undertaken in order to understand new phases and phase behaviors that are being observed in such systems.

One particular system that has received quite a bit of attention in recent years has been that of physisorbed layers of *n*-alkanes on different solid surfaces. The interfacial properties of *n*-alkanes appeal to many wide-spread applications and problems (e.g. lubrication, adhesion, catalysis, etc.) in addition to the further understanding that this physical system contributes to that of "similar" systems. Such similar systems include lipids (which contain an alkane in their hydrophobic tail) and more complex molecules such as some polymers, which have similar C-H bond arrangements to alkanes (e.g. polyethylene). In addition, understanding the behavior of *n*-alkanes at an interface directly relates to industrial problems as broad as petroleum flow to an understanding of the processes that take place in ice cream.[1]

There has been a vast amount of experimental effort[2-16] over the past several years to study the bulk and interfacial behavior of *n*-alkane layers. Clarke and coworkers[2-6] have studied solid layers of *n*-alkanes at the graphite interface with $5 \leq n \leq 15$. They propose that a solid layer of *n*-alkanes at the interface exhibits ordering above the bulk melting temperature for layers containing molecules with chains longer than pentane ($n=5$)[4], suggesting a role of chain length on ordering in the layer near the melting transition. In addition, Taub and coworkers[7-10] have studied the melting behavior in layers of tetracosane ($n=24$) and dotriacontane ($n=32$) on graphite through neutron scattering experiment. These authors report melting transitions of these monolayers well above the bulk melting transition, and neutron diffraction data suggests a melting transition for the tetracosane (C24) monolayer beginning at temperatures near ca. 340 K. These authors have also studied the structure of 1.94 layers of C24 on graphite, and have determined that at coverages greater than a complete monolayer, the bottom and top layer tend to arrange in positions that are fully commensurate with the underlying graphite substrate.

Other experimental work has been devoted to studying the behavior of alkanes of size comparable to those studied in this work, through methods of calorimetry[11-12] and scanning tunnelling microscropy (STM).[13-16] In particular, calorimetry experiments[11-12] suggest that bulk $n$-alkane crystals with alkane chain lengths between $n$=21 and $n$=32 undergo several rotator phases prior to melting. Additionally, STM studies provide a glimpse of how these alkane layers arrange at the graphite interface in the crystalline solid, prior to melting. Particularly relevant to this work is an STM study of octacosane ($C_{28}H_{58}$) and dotriacontane ($C_{32}H_{66}$) monolayers on graphite by Bucher et al.[14] These authors use STM to study the disordering of the solid phase, and suggest that fluctuations of the individual chains and groups (or "blocks") of chains evolve with increasing temperature until the lamellae in the solid monolayer are no longer distinguishable. These authors also observe increased conformational disorder (gauche defects) forming in the chains as well. The picture that is painted of the disordering process by Bucher et al. is well paralleled through the simulations conducted in this work.

In addition to experiment, there has been a significant amount of simulation studies[17-25] conducted over $n$-alkane layers on graphite. Simulations conducted by Hansen et al.[17-19] have focused on classifying the melting transition in monolayers of C24, and how this melting transition is indicative of the well-known gel-to-fluid transition observed in lipid bilayers.[17] The presence of intrachain melting and centralized gauche defects is proposed to create the space reduction necessary for the melting transition to take place. Although this work provides an excellent outline of an interesting phase transition that takes place in this monolayer, it does not go into depth in studying the solid phase of the monolayer at temperatures leading into the transition, as is the purpose of the study at hand.

Recent studies over monolayers of shorter chained alkanes have also brought some interesting observations. One such observation is that layers of $n$-alkanes of chain length as small as decane ($C_{10}H_{22}$) at monolayer coverage on graphite exhibit a phase transition that is accompanied by a significant number of gauche defects in the alkyl chains[20]. In addition, simulations of pentane, heptane, and nonane[21] indicate that monolayers involving alkanes with chains as small as heptane exhibit a disordering process that (i) occurs over a large temperature range (as compared to shorter alkanes), (ii) involves disordering that begins through the "shifting" of lamellae, and (iii) is observed for the two longer alkanes studied, independent of gauche defect formation.

In addition, it should be noted that there have only been a few simulation studies of *n*-alkane bilayers adsorbed on graphite. The first of these, conducted by Peters[22], suggests that melting for the hexane ($C_6H_{14}$) bilayer occurs ca. 20 K above the melting temperature of the monolayer, in agreement with experiment. In addition, Peters also suggests that the transitions in the bilayer mimic those of the monolayer, with the exception that the transitions in the bilayer are defined more broadly with respect to temperature. Additional simulations by Krishnan et al. have utilized an all-atom simulation model to study hexane monolayers and trilayers[23,24] as well as heptane monolayers and bilayers[25] on graphite. Observations of the melting behavior of the heptane bilayer provide interesting insight into a concept of layer-by-layer melting, which is a significant aspect of this work. These authors suggest that heptane bilayers are more stable in a fully commensurate solid phase, and that distinct features of the energetics in the monolayer and bilayer suggests different melting behavior in each layer. This observation of the melting transition based upon layer-by-layer melting is the foundation for the study at hand, which investigates the melting in much longer alkanes than have been studied in previous experiment, at coverages greater than a monolayer.

Thus, it is the purpose of this work to better elucidate the melting transition in the tetracosane monolayer through simulations, utilizing bigger simulation cells and a thorough analysis of the structural and molecular behavior as compared to previous work over C24 monolayers. In addition, this work is conducted to investigate the melting of a bilayer of a longer-chained alkane in comparison to a monolayer. There have been no investigations regarding the *mechanism* of melting in molecular layers with coverages deviating from the strict single-layer coverage studied in most previous work, nor have there been investigations of the melting in bilayers of alkanes with chain lengths greater than heptane ($C_7H_{16}$). Thus, there is a plethora of important questions surrounding this system which deserve to be investigated. In particular, the understanding of how melting proceeds in the molecular layers as the coverage approaches the bulk is of great interest. Many phenomena, including surface freezing[26], have been recently discovered and still reside as being unexplained. The work presented here attempts to provide a link between ideas about melting proposed for *n*-alkane monolayers on graphite (i.e. the footprint mechanism) to a bilayer, to better understand how the mechanism of melting evolves with coverage.

## 2 Simulation Details

All simulations reported in this study are performed through use of an isothermal canonical ensemble molecular dynamics (MD) method. The potential model and isothermal temperature control that is utilized in simulations conducted in this work is the general model used in most similar simulations, and is described in detail in previous work.[20,27] Thus, only a brief description will be given here, and the reader is referred to previous published literature for a more detailed description.

The total simulated potential energy is a function of both non-bonded and bonded potential terms. The non-bonded terms involve atom-atom interactions (modeled by a 12-6 Lennard Jones interaction potential) and adsorbate-graphite interactions (modeled by Steele's graphite interaction potential[28]). In addition, the bonded interactions encompass three-body and four-body intrachain bending terms with potentials describing these bond-angle[29] and torsional[30] motions that are consistent with previous work.

In order to simulate the alkane molecules, the most recent anisotropic united-atom (AUA4) model[31] is utilized which approximates each methyl and methylene group as a single (psuedo) atom, with the mass of the methyl or methylene group. In addition, this model provides interaction parameters for the Lennard-Jones potential utilized which slightly differ between the two groups, giving a more realistic physical interpretation of the interactions between atoms in an alkane molecule.

Table I contains information regarding simulation cell sizes, lattice parameters, and quantities of molecules studied in simulations. To perform simulation of the solid-liquid phase transition, simulations are began from a low-temperature solid phase with lattice parameters determined through diffraction experiment. The cell is then allowed to equilibrate to a configuration through a period of time that allows the layer(s) to come to thermodynamic equilibrium. For the case of the monolayer, the equilibration period spans for 750 ps (7 x $10^5$ steps), followed by a period of 450 ps in which averages are taken and thermodynamic quantities are calculated (total simulation time of 1.2 ns). In the case of the bilayer, the equilbration period is taken to be for the first 700 ps, and averages are taken over an additional 700 ps (total simulation time of 1.4 ns). With a cell size involving 2304 atoms and cutoff radius of 10Å for adsorbate-adsorbate interactions, each simulation took ca. 780 hours on the UNI Opteron Cluster, which is an X86-64 Linux cluster. A total of 38 temperature points are sampled for the bilayer, and 33 temperature points are sampled for the monolayer- in adddition to some variations conducted for the monolayer. Variations are typically run for a period of 1 ns, with equilibration carried out for the first 500 ps of simulation time.

## 3  Results

### 3.1 Structural Characterization of Melting

Over the past several decades, many interesting observations have been made for adsorbed atoms and molecules on a solid substrate. These observations tend to all originate from the very interesting balance between interlayer interactions and layer-substrate interactions, which tends to lead to a large number of interesting phases and effects in the adsorbed layer. In this study, the first step to understand the properties of the C24 thin films on graphite is to look at characteristics of the molecular structure in order to analyze how melting takes place.

The first quantity that one may be interested in when looking at melting from a structural perspective is a structural order parameter that exploits the symmetry of the adsorbed solid phase. In this study, the structural order parameter, $OP_2$, is shown in Fig. 1 and defined as follows:

$$OP_2 = \frac{1}{N_m}\left\langle \sum_{i=1}^{N_m} \cos(2\phi_i) \right\rangle$$

where $\phi$ is the in-plane azimuthal angle of molecule $i$.  At low temperatures, the C24 molecular long axes in the solid phase assume an orientation that can be easily monitored through use of this order parameter, and hence this quantity gives excellent insight into the (i) solid phase behavior and (ii) melting behavior of the C24 monolayer and bilayer.  In addition, $OP_2$ for the bilayer is averaged seperately for the bottom and top layers.  The criteria for separating these two layers is that the molecular centers of the molecules in the top layer be greater 1.5 Å less than the average height of all molecules in the top layer.  This is found to be a good criterion, and is utilized throughout the remainder of this work where averages are taken over the bottom and top layers of the C24 bilayer seperately.

      Analysis of Fig. 1 brings forth a few initial observations that support some of the main features of melting observed in this study.  First of all, the most apparent feature in Fig. 1 is the differing values of $OP_2$, as well as the different melting temperatures predicted by $OP_2$, for C24 monolayers and bilayers.  One can observe that, after ca. $T$=340 K, the C24 monolayer starts to evolve out of the ordered phase, and by $T$=360 K, the monolayer has essentially disordered, even though some short-range order still exists.  For the bilayer, one can observe that the melting process for the bottom layer is aesthetically (with respect to $OP_2$) similar to that of the monolayer, except the molecular orientations are better aligned with the maxima in $OP_2$ for the case of the bilayer.  However, the top layer (in the C24 bilayer) indicates different melting behavior than the bottom (interfacial) layer.  The loss of strict rectangular-centered (RC) order in the top layer begins at temperatures above 355 K, whereas the loss of strict RC order in the bottom layer begins at those above 375 K- a 20 K difference!  However, in the temperature range of 355 K to 375 K, the top layer still seems to exhibit a significant amount of order- which is probably induced by the underlying C24 layer.  However, as soon as the bottom layer starts to lose the strict RC order, the melting transition seems to proceed quite rapidly, and by 390 K, the bottom and top layers have both evolved into a fluid, with the top layer surprisingly exhibiting a bit more short range order than the interfacial layer.

      One additional feature of Fig.1 that is very peculiar is the differing values of $OP_2$ for the monolayer and bilayer at low temperatures- indicating slightly different arrangements in the solid phase.  This can be better understood through Fig. 2, where the distributions of the in-plane azimuthal angles,

averaged over all post-equilibrated simulation time, are presented for several temperature points before and near the melting transition.  For the monolayer at low temperatures, one can observe that two distinct peaks seem to arise- indicating a slight rotation of the molecules.  At low temperatures, the molecular orientations seem to take on values of ca. 90° ± 20°, which then evolves to orientations with rotations closer to 90° until temperatures near 330 K, where the orientations seem to have a single broad peak around 90°.  This continues until melting, and orientational disorder follows.  For the case of the bilayer, the molecular orientations seem to be primarily peaked at 90° (resulting in a value of $OP_2$ very close to 1), with two smaller peaks at ca. ± 10° on either side of the large peak.  This generally continues throughout melting as the peaks evolve toward broadening about the large peak at 90°.  To visually represent the orientations described here, snapshots are provided in Fig. 3 that represent (i) the low-temperature monolayer, (ii) the monolayer prior to melting (where the peak in the orientations at 90° develops), and (iii) the low-temperature bilayer.  It should be noted that comparisons between the top and bottom layers in the C24 bilayer indicated no difference in the structure- with the exception that the bilayer tends to evolve toward disordering at lower temperatures.

In addition to the in-plane structural characterization, it is also useful to understand the layering structure- which is illuminated by the atomic height profiles presented in Fig. 4.  For temperatures before and near melting for the monolayer, the top panel of Fig. 4 indicates a broadening of the height profile, accompanied by the formation of a small shoulder as the temperature is increased.  For the C24 bilayer, however, the observation of a broader distribution in the top layer than the bottom layer is a unique feature to this system.  Although this was observed in a previous study of heptane bilayers on graphite[25], it seems to be more pronounced for the C24 bilayer.

**3.2 Molecular Behavior Before and Near Melting**

It has been demonstrated in previous simulations[20] that melting in layers of chain molecules with significant chain lengths is a process dominated by the molecular behavior and gauche defects forming in the chains. In particular, it is reasonable to wonder how these chain defects and molecular behavior play a role in melting in the bilayer, as compared to the monolayer.

In order to study gauche defect formation in the C24 chains, the average methyl-methyl distance is calculated and averaged over the post equilibrated simulation time, and presented in Fig. 5. At low temperatures, the methyl-methyl distance in the C24 monolayer and bilayer takes on a value of ca. 29 Å, which represents the solid layer with all molecules taking on an all-trans conformation. As the molecular chains start to kink due to gauche defect formation, the methyl-methyl distance in the chains starts to decrease in accordance with the type of "kink" or gauche defect in the chain. In particular, one can categorize the gauche defect formation, through this quantity, into three different regions. For the C24 monolayer, the following three regions are evident: (i) trans to end-gauche conformation at temperatures between 200 K and ca. 320 K (minimal deviations from the all-trans molecular conformations), (ii) central-gauche conformation at temperatures between ca. 320 K and 350 K (significant deviation from all-trans molecular conformations), and (iii) intrachain melting at temperatures above ca. 350 K (convergence of the methyl-methyl distance to values comparable to minimum-energy atomic pair distances). Analyzing the bottom panel of Fig. 5, it is evident that these three regions are observed for the top and bottom layers of the C24 bilayer, except that the temperatures over which the regions are defined for the bilayer are significantly shifted from the case of the monolayer. In particular, the top layer undergoes a central-gauche region at temperatures ranging from ca. 290 K to 360 K, and the bottom layer seems to exhibit this region over the temperature range of ca. 355 K to 375 K. This shows evidence of layer-by-layer melting based upon the individual molecular behaviors, which will be discussed in further detail in the next section.

In addition to gauche defect formation, previous work over heptane monolayers and bilayers[25] has shown that there exists a very interesting coupling of gauche defect formation and molecular rolling that plays an important role to preempt the melting transition. Thus, in this study, a bond-roll order parameter, $\beta_{roll}$, is developed in order to elucidate the molecular rolling behavior before and near the melting transition. This parameter is defined as:

$$\beta_{roll} = \frac{1}{N_m(N_a-2)} \sum_{i=1}^{N_m} \sum_{j=2}^{N_a-1} \frac{(\vec{r}_1 \times \vec{r}_2) \cdot \hat{z}}{|\vec{r}_1 \times \vec{r}_2|}$$

where the index $j$ is over the total number of psuedoatoms in a molecule, and the index $i$ is over all molecules. The vectors, $r_1$ and $r_2$, correspond to bond vectors that are directed toward the ($j$+1)th and the ($j$-1)th psuedoatoms in the molecule chain. This quantity takes on a value of unity if all molecules lie on the substrate with their bond vectors parallel the the substrate plane, and disappears when the molecules are "rolled," or are arranged with their molecular backbones rigid with the substrate. Analyis of Fig. 6 for the C24 monolayer indicates that the rolling of the molecules in the layer is a temperature-induced effect. At ca. 200 K for the monolayer, there is a small amount of molecular rolling that is observed, which increases almost linearly until temperatures after the region where the central-gauche defects are suggested to form (from Fig. 5). Interestingly, the temperature dependence of the molecular rolling in the bilayer seems to be analagous to that observed in the monolayer, except there seems to be a greater number of molecules in the bilayer rolled at lower temperatures. Conversely, the bottom layer of the C24 bilayer is the only layer observed to exhibit (almost) completely flat molecular orientations at low temperatures. Apparently, there seems to be a coupling effect of the increasing temperature and both the formation of gauche defects and onset of molecular rolling. This leads to several avenues of thought as to the mechanism of how the melting transition *actually* takes place, however, one way in which a better understanding of this may emerge is through an understanding of the fluctuations in the layer, studied in the next section.

### 3.3 Molecular Fluctuations Prior to Melting

At the heart of phase transitions in atomic and molecular layers is the idea of fluctuation-induced melting. As early as the 1960's, Mermin and Wagner[32] proved that long-ranged positional order in a 2D layer can not truly exist due to the role of long-range fluctuations. Following this, in the 1970's, a series of contributions from Kosterlitz, Thouless, Halperin, Nelson, and Young (KTHNY)[33] provided insight into a two-stage, dislocation mediated melting transition for a layer in two dimensions. Recent theories of melting for layers of chain molecules has illustrated a qualitative mechanism for melting based on out-of-plane behavior[34], and has speculated that out-of-plane molecular behavior (such as gauche defect formation) tends to "facilitate the rotational and translational motion needed" for the melting transition to take place.[7] Thus, there is still no rigorous understanding of the melting transition in these layers, or how fluctuations in the layer contribute to melting, as well as other effects such as gauche defects.

The first quantity studied, and presented in Fig. 7 is the total mean square out-of-plane fluctuations, defined as the square difference in the center-of-molecule out-of-plane position over all time, and the out-of-plane position at the first step after the equilibration period- averaged over all C24 molecules and all time steps, for the C24 monolayer and bilayer. It should be noted that this quantity was calculated by studying the fluctuations of the molecular centers, and not the centers of mass. In particular, a study of the molecular out-of-plane fluctuations through molecular centers is not distorted by the shift in the center of mass that takes place when a gauche defect forms in the molecule. This gives better insight into the *molecular* fluctuations, and does not allow the molecular conformation changes to be characterized as fluctuations.

In both layers, at temperature points in the solid phase, it is clear that the fluctuations increase with temperature in a rather linear fashion. However, near the melting temperature, both plots indicate out-of-plane fluctuations that break the linear trend established at lower temperatures and exhibit very significant amounts of out-of-plane behavior. However, for the bilayer, it is evident that the out-of-plane fluctuations start to emerge prior to the bilayer melting temperature, which may have significant relevance to the evolution of 2D melting to bulk melting, as well as the observed 3D rotator phases in the bulk. However, this observation suggests that out-of-plane behavior *does* facilitate the melting transition in both the monolayer and the bilayer (but maybe not in exactly the same way).

In addition to the out-of-plane fluctuations, the fluctuations of the molecules in the layers are analyzed through the molecular center density profiles, calculated in the direction of the molecular long axes and presented in Figs. 8a and 8b. Since these profiles are averaged over the post-equilibration simulation time period, it gives a good indication of how the fluctuations in the layer evolve as the temperature is increased toward melting. Since the molecules arrange in lamellae in the solid phase, a very distinct peak corresponds to a lamellae that has all molecular centers aligned. However, with the onset of fluctuations in the layer, the molecules will start to undulate within the lamellae, and the distributions will broaden. In a fully disordered liquid, the distributions will not have any distinct features, and the lamellae will no longer be distiguishable. One interesting feature of Fig. 8a is that the picture that is illustrated by STM studies of similar alkane layers[14] is evident through these density profiles. At low temperatures, the molecules primarily reside in fixed lamellae, with some fluctuations in the layer, but generally with well-defined positions. However, upon increasing the temperature to ca. 300 K, one notices that there is a very significant broadening of the density peaks, corresponding to significant fluctuations about the molecular long axes. At temperatures of 340 K, near where the layer undergoes melting, one can notice that the density distributions are no longer uniquely defined, and the density peaks start to merge together. This is evident of the process defined by Bucher et al. where the lamellae are observed to lose their individuality near the melting transition[14]. In the case of the bilayer (Fig. 8b), it is evident that at low temperatures, the density distributions of the bottom layer are very fixed, suggesting very little fluctuations of the molecules at the interface. However, the top layer seems to exhibit more broadening about the central positions of the lamellae, indicating more fluctuations in the top layer than the bottom layer. This is true at all temperature points presented in Fig. 8b, which is indicative of the layer-by-layer behavior exhibited in other quantities. In particular, this would suggest that the increase in the melting temperature of the bottom layer in the C24 bilayer (with respect to the monolayer) may be due to a "confinement" effect, where fluctuations are hindered by the presence of a top layer. Such an idea may provide useful to better understand how fluctuations and intramolecular defects work together to play a role in the melting process. This will be further detailed through a series of variations provided in the next section.

### 3.4 Variations

Utilizing simulations to understand experimental observations allows one to change specific features of the simulation and the system under study to a hypothetical scenario, and observe how the system in this scenario proceeds through the process being studied. This route has been consistently utilized in previous work to understand how gauche defects play a role in the monolayer melting process. This section will outline four different variations that were conducted for the C24 monolayer, each for a better understanding of the link between gauche defect formation, intermolecular packing of the molecules in the solid phase, and fluctuations of the molecules in the solid prior to melting.

The first variation that is carried out is the simulation of the C24 monolayer with the constants defining the series in the dihedral potential increased by an order of magnitude. Such a method was employed in previous work to understand how gauche defects play a role in melting. However, in the present study, this is carried out in order to understand how fluctuations in the monolayer are coupled with gauche defect formation. From the snapshot presented in Fig. 9, it is evident that at 300 K, there is a significant amount of lamellar fluctuations in the monolayer. However, melting is not observed to occur at temperatures lower than 475 K, and one can immediately observe that gauche defect formation plays a huge role in this transition. This suggests that, in such alkane layers, the melting transition proceeds through a two-step process, with the first step being the fluctuation of the molecules in the layer about their long axes, and the second step being the formation of space from gauche defect formation. However, independent of gauche defect formation, there still exists the lamellar fluctuations observed for the layer with gauche defects, indicating that it is not particularly the gauche defects which are responsible for the lamellar disorder, but possibly vice versa. In order to better understand this idea, the monolayer was simulated with the same number of molecules, but the graphite substrate was made large enough that no molecule was close enough to another molecule to interact, and thus the only dominating interaction in this configuration is the graphite-adsorbate interaction. Interestingly, it is observed that at temperatures well below the monolayer melting temperature, the molecules are dominated by gauche defect formation, and primarily central gauche defect formation. This suggests that the molecules interacting with the graphite alone, will tend to undergo chain kinks at temperatures well below the monolayer melting, due to the graphite-adsorbate interaction. It is well known that in the solid phase, it takes much more energy to initiate out-of-plane motion, because the molecules are "held" to the substrate by the strong holding potential, as well as being closely packed within lamellae, which supports the restriction of in-plane motion. However, when there is no in-plane restriction of the molecular motions, it is very easy for the gauche defects to form in the molecule chains. This suggests that the packing arrangement of the adsorbed monolayer plays a significant role in that it imposes the condition that the molecules can only disorder via intrachain kinks through either (i) out-of-plane motion (which is highly energetically unfavorable at low temperatures), or (ii) shifting of the molecules away from the aligned lamellae so that their chains can have in-plane freedom to form gauche defects. This idea was speculated in previous work by Hansen[18] as well, even though these variations confirm this picture.

In addition, to better understand the role that intermolecular packing plays on the monolayer melting behavior, the C24 monolayer is modeled as being fully commensurate- with the monolayer having the same lattice parameters as the fully commensurate bilayer observed in previous experiment. Experiment has suggested that the C24 molecules prefer to arrange in a uniaxially incommensurate layer, and the simulations carried out with this configuration indicate that the total energy of the fully commensurate monolayer, at all temperature points, is higher than that of the incommensurate monolayer reported in the previous section. However, one interesting feature of this system is that the fully commensurate monolayer melts at a temperature that is higher than the uniaxially incommensurate monolayer- at ca. 380 K (comparable to the bottom layer melting temperature of the bilayer). Not only does this underline the very delicate nature of the long-chained alkane/graphite surface phase diagram (set by the coverage and commensurability), but it also gives very interesting insight into the role that the surface interaction plays in the monolayer melting transition. In particular, the role of the surface interaction can be illuminated by comparing the fully commensurate layer to the uniaxially incommensurate layer. In the fully commensurate layer, where the molecules minimize their interaction energy with the underlying substrate, the melting transition temperature occurs significantly higher than the melting transition for the uniaxially commensurate monolayer, where the intermolecular interaction plays a bigger role in melting (and melting still occurs at a temperature higher than the bulk melting temperature).

Finally, the last variation that was considered was to analyze the melting behavior of the monolayer when the pair interactions were not taken beyond first neighbors (~ 5 Å). Although it is well-known that a decreased interaction energy will lead to a lower melting transition (which is responsible for the lower bulk melting point of branched alkanes with respect to linear alkanes), it is not quite understood in this context the role of the intermolecular interactions on the 2D layer. In the right side panel of Fig. 9, a snapshot of the C24 layer at 300 K is presented, where the pair interactions were truncated as described above. At this temperature, one can already notice the in-plane interdiffusion of the lamellae as well as the significant gauche defect formation, and simulations suggest that melting in the layer occurs at ca. 310 K, about 30-40K lower than that of the monolayer with pair interactions defined out to 10 Å. This observation compliments that of previous variations conducted over gauche defect formation in the chains with non-interacting molecules- further suggesting the role of the intermolecular interactions in melting in the quasi-2D layer.

## 4 Discussion

### 4.1 Layer-by-layer melting

Layer-by-layer melting of an adsorbed bilayer on graphite was an effect that was reported in a recent MD study of heptane bilayers on graphite by Krishnan et al.[25] Such an observation has significant relevance toward an understanding of the surface freezing effect that has been observed and studied for multilayers of alkanes on graphite.

One particular observation in this study of C24 on graphite is evident through a comparison of Fig. 1 and Fig. 5. In Fig. 1, one can observe that, until ca. 20 K below the bilayer melting temperature, the structural characteristics of the top and bottom layers are not distinguishable, however, their molecular behavior is quite different. In addition, it is evident that, even though the value of $OP_2$ in Fig. 1 for the top layer starts to decrease from the value of $OP_2$ in the bottom layer, it still does not achieve a value of $OP_2$ less than 0.8, which still implies significant ordering in the top layer, despite the greater abundance of gauche defects in the C24 chains in the top layer. If one considers this disordering process in terms of the footprint reduction mechanism proposed by Hansen et al.[34], then one may be preemptive to conclude that the top layer "melts" before the bottom layer (due to the space reduction created by the C24 molecules with gauche defects in the top layer). However, until the bottom layer melts, the value of $OP_2$ for the top layer does not achieve a value less than 0.8, suggesting an imposed order in the top layer resulting from the solid C24 layer at the graphite interface. Based on this observation, it seems that the behavior of these two layers is coupled. It does seem as if the top layer of the C24 bilayer *wants* to disorder first, but it can not exist in a "disordered" state without the bottom layer first undergoing the order-disorder transition.

In addition, if one considers the out-of-plane fluctuations present in the bilayer compared to the monolayer shown in Fig. 7, one notices that, for the monolayer, the out-of-plane fluctuations begin to break the linear trend *at* the melting transition. However, for the bilayer, the out-of-plane fluctuations seem to break this linear dependence at the point where the top layer seems to lose it's strict RC order (and not the melting transition). Such an observation evidences that the difference in melting between two molecular layers and one molecular layer is that only out-of-plane motions contribute to the melting in one layer, while the melting process in two layers is significantly more complicated- involving fluctuations and space reduction in the top layer that does not result in the melting transition.

In light of recent experimental[26] and theoretical[35] work which has illuminated the surface freezing phenomenon in such alkane layers and the relation of molecular weight and fluctuations to this phenomenon, it is interesting to consider the results of this study to understand the mechanism of this effect. One observation of this study is that, even amidst out-of-plane fluctuations and gauche defect formation of the molecules in the top layer, the bottom layer will still exhibit good RC order, with very little gauche defect formation in the molecules until temperatures near melting. Although three C24 layers are not considered here, the author speculates that further progression toward the bulk C24 would induce out-of-plane rotator phases that will play a large role in the melting away from the interface. However, it is the "confinement" of the bottom layer that seems to play a role in the melting behavior. In a multilayer system, the first few layers will have a significant strain introduced in the layer from the bulk 3D behavior near the melting transition. This should create a strain energy at the interface where it would take more thermal energy to overcome the decreased ability of the molecules to undergo out-of-plane motions, and hence undergo melting. Such an idea is complimented by simulations that have previously reported melting temperatures for alkane monolayers on graphite that have been above the bulk melting temperature. Although simulations of many layers of alkanes is still computationally impossible, the progression from one to two layers of C24 at the interface may suggest trends that are evident at higher coverages.

### 4.2 Comparison to Experiment

The results of this work seem to be in good agreement with experimental observation. Previous neutron diffraction experiment conducted over C24 monolayers[7-8] suggests that the diffraction peaks start to broaden at ca. 340 K, and continue to broaden and shift through ca. 350 K. The structural observation in this work of the loss of solid phase order beginning at 340 K, proceeding through a disordering process that involves complete disorder at ca. 360 K (aided by molecular chain melting) is in excellent agreement with experimental diffraction data. Previous simulations have also been conducted over the C24 monolayer[18], where the melting temperature was determined to be at 338 K, even though it was not discussed by the authors how "melting" was defined. This is important since the C24 monolayer does not exhibit an extremely well defined (or sharp) transition with increasing temperature, but rather involves a process that is gradual with respect to temperature due to the imposed order that exists after melting from the long molecular chains. Upon inspection of the temperature-dependent neutron diffraction patterns presented in ref. [7], one can observe that there is only a very slight shift in the main peak (in addition to some broadening of the peak) between ca. 345-351 K. In general, this would suggest a gradual melting transition and residual order in the temperature region of 340-350K that seems to match the observations in this work. Although there is no direct experimental comparison of melting temperatures for the bilayer, the increased melting temperature of the bilayer with respect to the monolayer was an effect that was also observed in experiment[36] and simulations[22] for hexane on graphite, and it is not unreasonable for this to be the case for the C24 bilayer as well.

In addition, previous STM work by Bucher et al.[14] gives a fascinating insight into a visual depiction of the melting process in the C28 and C32 layers, which are very similar to the C24 layer studied in this work. These authors report a process that involves longitudinal motion of the lamellae that make up the layer until a temperature where the lamellae lose their individual identity. Such an observation corresponds well to what one would expect, given observations in simulations of shorter alkanes[21], where similar behavior is observed and attributed to the alkane chain length. In addition, the simulations conducted here seem to show very similar behavior to that observed through STM studies of the C24 monolayer, which is evident from the density distributions in Fig. 8a. However, Fig. 9 shows that the longitudinal motion of the C24 molecules in the layer is still evident when there are no gauche defects in the layer- suggesting that the thermal energy plays a large role in these fluctuations. However, as one might expect, melting does not occur in such simulations until very high temperatures, indicating that gauche defects play the main role in the melting transition. In addition, Fig. 9 also indicates that a smaller interaction radius (and hence, interaction energy) promotes a significantly lower temperature (prior to chain melting), indicating that the interaction energy plays a large role in sustaining order above the bulk C24 melting temperature until melting. However, analysis of Fig. 6 also indicates that there is significantly less longitudinal motion of the molecules in the bilayer, as opposed to the monolayer, which could result from the 3D interactions between molecules of the two layers resulting in a "confinement effect" of these motions. However, the observations in this work seem to be in good agreement with those of Bucher et al.

Finally, it should be noted that the observation of Krishnan et al.[25] of the significant role in perpindicular rolling of molecules near the melting transition is apparent in this study of C24 monolayers and bilayers as well. One factor that seems to preempt the phase transition is molecular rolling, even though it seems that there is significantly more rolling for the top layer of the bilayer than either the monolayer or the bottom layer of the bilayer, as shown in Fig. 6. This seems to be consistent with experiment, as Matthies shows a good fit to the 1.8 monolayer C24 diffraction pattern with an alternating parallel/perpendicular orientation of the molecules in the bottom/top layers, respectively. In general, simulations suggest that the bottom layer seems to involve more molecules that are parallel to the surface than the monolayer as well.

## 5 Conclusions

This work gives insight into the melting behavior of tetracosane monolayers and bilayers adsorbed onto a graphite substrate. Through analysis of the melting transition and the solid phase behavior before the melting transition, simulations suggest some interesting features of melting in the bilayer as opposed to the monolayer. First of all, a layer-by-layer melting effect is observed, even though the disordering of the top layer does not ensue until the bottom layer loses order as well. This suggests that the idea of a space reduction does not explicitly apply to a bilayer, but one can possibly think of the partial disordering of the top layer before the bottom layer as a two-stage space reduction, where vacancies must occur in the top layer and within the bottom layer for melting in the bottom layer to take place. In addition, significant features linked to melting, such as molecular rolling and gauche defects, are observed to occur in the top layer of the C24 bilayer before the bottom layer. The results of this work seem to be in good agreement with diffraction data of C24 on graphite, as well as STM results and images of similar long-chained alkanes on graphite. This work, in addition to previous simulation studies of heptane bilayers on graphite, gives a better understanding of melting at an atomic level for both shorter and longer alkanes on graphite, which places a framework for a complete picture of melting as the interfacial layers exceed a single molecular layer of coverage.

**Acknowledgements**

The author is grateful to Paul Gray and the UNI CNS for use of the UNI Opteron Cluster to carry out calculations for this work. Also, the author is grateful to John Deisz and the UNI Physics Department for support and additional computing resources.

**Tables and Figures**

**Table I**. Simulated cell sizes, lattice parameters, and total number of simulated molecules for the tetracosane (C24) monolayer and bilayer.

|  | C24 Monolayer | C24 Bilayer |
|---|---|---|
| cell dimensions (Å) | 72.48 x 129.2 | 51.12 x 130.6 |
| $a, b$ (Å) | 4.53, 64.6 | 4.26, 65.3 |
| $N_m$ | 64 | 96 |

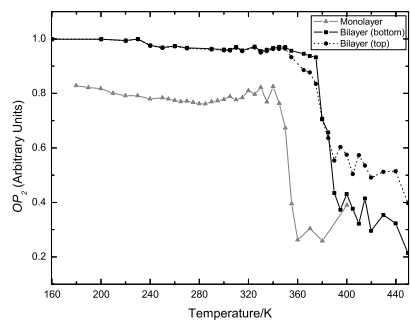

**Fig. 1**. Structural order parameter, $OP_2$ for monolayers and bilayers of C24. Averages are taken over the top and bottom C24 layers independently.

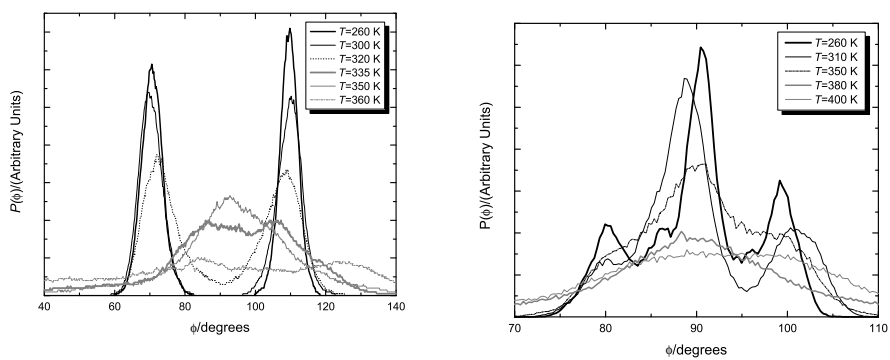

**Fig. 2** Orientation profiles for the C24 monolayer (top panel) and bilayer (bottom panel) for temperature points near and after the melting transition in both cases.

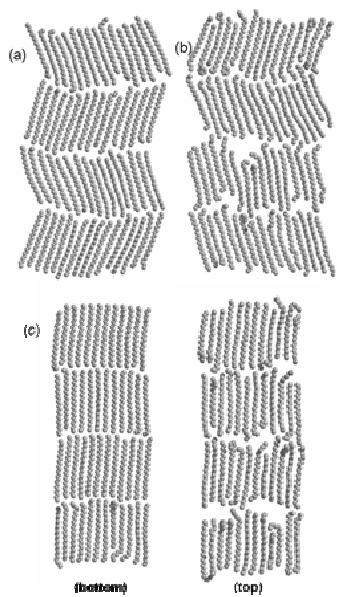

**Fig. 3.** Snapshots of (a) the low-temperature crystalline monolayer C24 solid phase in simulations (260 K), (b) the C24 monolayer at temperatures approaching melting (310 K), and (c) the top and bottom layers of the C24 bilayer at 320 K.

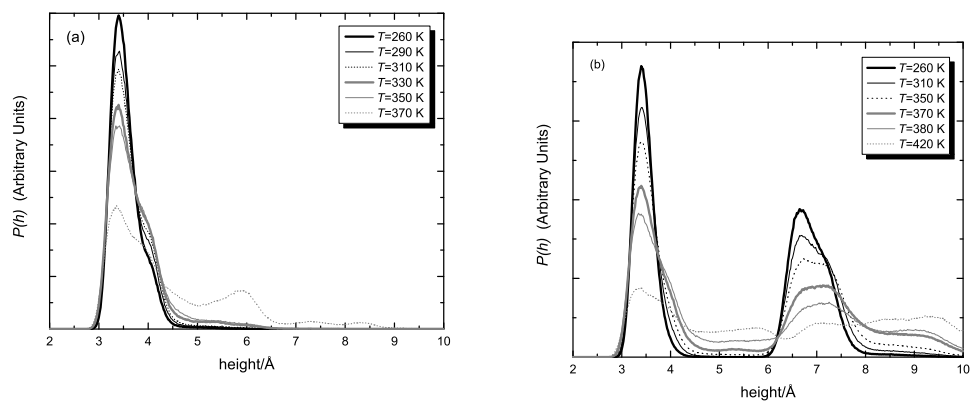

**Fig. 4.** Height profiles for the (a) C24 monolayer and (b) C24 bilayer at temperature points near and after the melting transition.

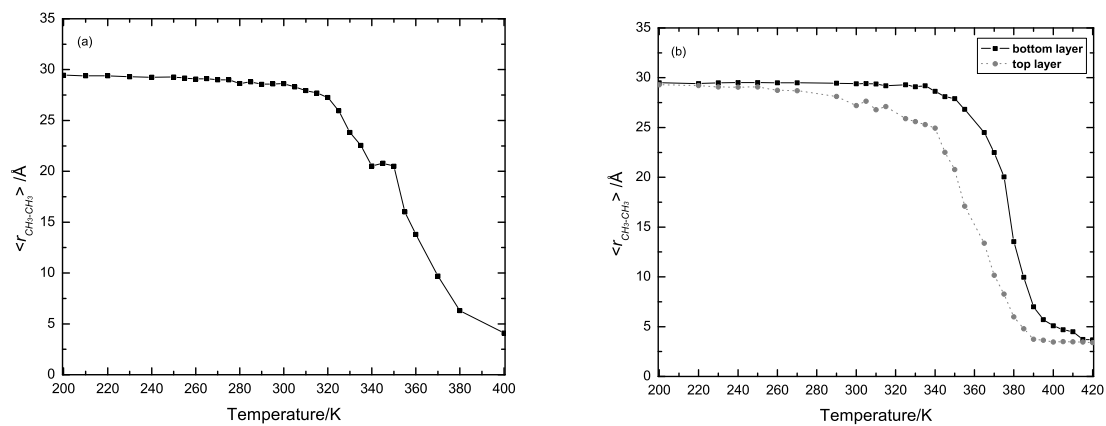

**Fig. 5** Average methyl-methyl distance at temperature points near and after the melting transition for (a) the C24 monolayer, and (b) the C24 bilayer, with averages taken over the top and bottom layers separately.

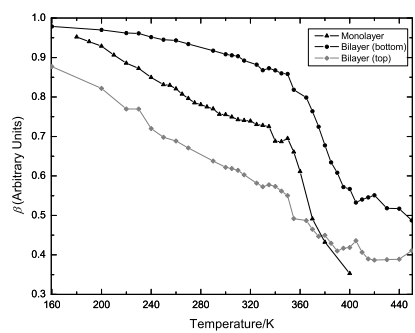

**Fig. 6.** Bond-roll order parameter, $β_{roll}$, for the C24 monolayer and bilayer at simulated temperature points.

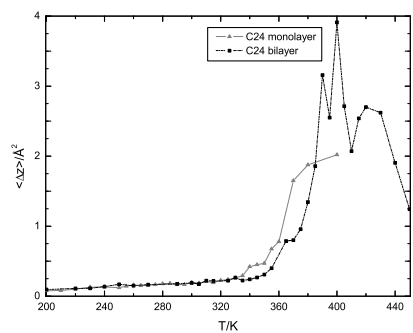

**Fig. 7.** Total mean square out-of-plane fluctuations at temperatures near and after the melting transition for the C24 monolayer and bilayer.

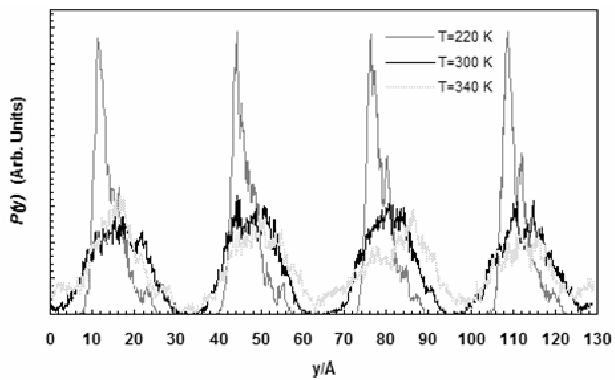

**Fig. 8a.** Center-of-molecule density distributions for the C24 monolayer in the direction of the molecule long axes, defined at labelled temperature points

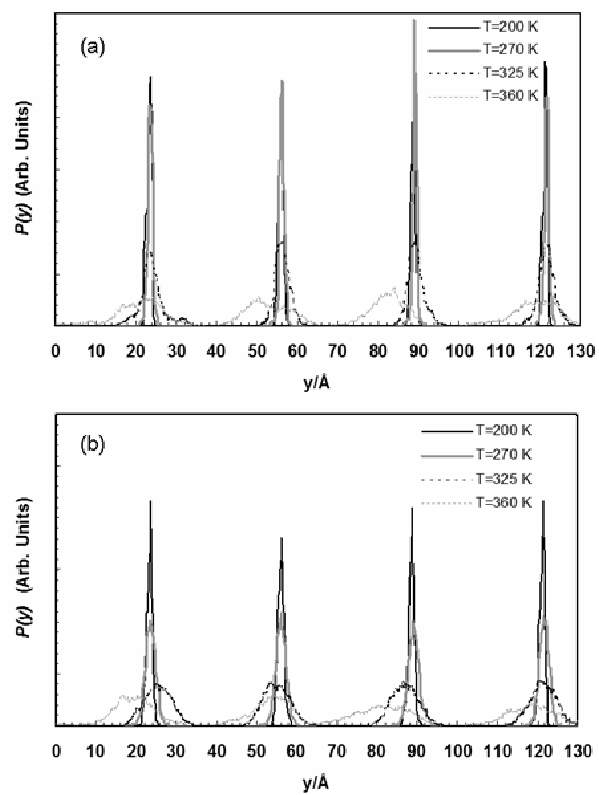

**Fig. 8b.** Center-of-molecule density distributions for the C24 bilayer, and specified temperature points. Distributions for (a) the bottom layer, and (b) the top layer.

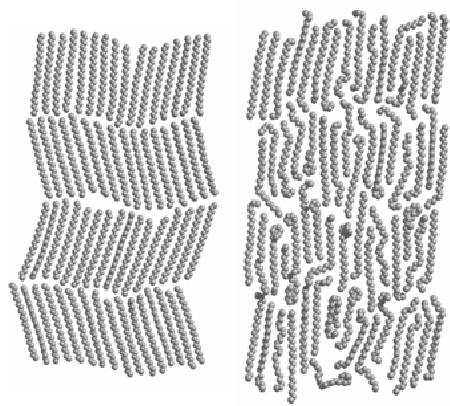

**Fig. 9.** Snapshots of the C24 monolayer at 300 K with (left panel) no gauche defects present in the layer, and (right panel) with only nearest-neighbor pair interactions allowed.